\documentclass[11pt]{article}

\usepackage[margin=1in]{geometry}
\usepackage[utf8]{inputenc}
\usepackage[T1]{fontenc}
\usepackage{amsmath,amssymb}
\usepackage{booktabs}
\usepackage{multirow}
\usepackage{graphicx}
\usepackage{xcolor}
\usepackage{tikz}
\usetikzlibrary{arrows.meta,positioning}
\usepackage[colorlinks=true,linkcolor=blue!60!black,citecolor=blue!60!black,urlcolor=blue!60!black]{hyperref}
\usepackage[round]{natbib}

\newcommand{\corpusA}{\textsc{Corpus~A}}
\newcommand{\corpusB}{\textsc{Corpus~B}}

\newcommand{\labelfmt}[1]{\texttt{#1}}

\title{Historical Backtesting for Scientific Question Discovery:\\
A Protocol and Astronomy Pilot\\[0.6em]
\large Evaluating AI-Generated Scientific Questions Against Future
Scientific Progress}

\author{%
  Hui Mao\\
  Independent Researcher\\
  hui.mao@alumni.upenn.edu
}
\date{August 17, 2026}

\begin{document}

\maketitle

\begin{abstract}
Systems that generate scientific research questions are currently
evaluated by expert scores, LLM-as-judge ratings, or curated case
studies---all subjective, none falsifiable. We propose a different
standard: \emph{future scientific engagement as an observable,
falsifiable proxy for one important dimension of a question's value}.
We formalize \textbf{historical
backtesting} as an evaluation protocol for scientific question
discovery: a system generates questions from a corpus frozen at a
historical cutoff, the questions are frozen before any access to later
literature, and a temporally isolated future corpus determines---via
fixed retrieval, a citation-constrained judge, and a declared
adjudication tier---whether each question was subsequently answered,
partially addressed, independently posed, or ignored, and whether its
underlying premise was supported or refuted. The protocol is
model-agnostic: any system that emits frozen questions can be scored,
and all metrics are defined independently of how questions are produced.
We release a reproducible astronomy benchmark instance
(cutoff 2020-12-31): a 2{,}512-paper past corpus, a temporally isolated
1{,}891-paper future corpus (2021--2026), frozen questions, retrieval
records, adjudicated outcome labels, and one-command metric computation,
plus a submission interface and four reference baselines evaluated
through the identical pipeline. In an initial set of ten questions
generated by an evidence-graph system from pre-cutoff literature only,
all ten were substantively engaged by later literature: two were
answered, seven partially addressed, one independently posed and still
open---and one question's underlying premise (a strongly subsolar water
abundance for HD~209458\,b) was subsequently refuted by three
independent analyses, the exact convergence the question called for.
A scaled second instance (astronomy v1L: 424 frozen baseline questions
against a 5{,}754-paper future corpus) then stress-tests the
small-sample conclusions and revises two of them: engagement rates
\emph{do} discriminate at $n=125$ (random 73\% vs.\ direct-LLM 96\%,
$p<10^{-4}$), and premise refutation is rare but not unique---chasing
highly cited results catches refutations on 3.2\% of questions, while
random templates stay at zero and acquire a measurable
11\%-answered floor. Finally, we turn the benchmark's deepest threat---LLM weights that have
read the future---into its subject: a generator decomposition (LLM-only
vs.\ deterministic evidence-structure vs.\ structure-plus-LLM
verbalization) crossed with a four-cutoff temporal stress test
(2010--2024, 798 judged questions) whose last window postdates the
model's training. LLM-only generation shows \emph{memorized relevance
without specific foresight}: near-ceiling engagement and the closest
phrasing to future literature at every cutoff, but an answered rate
flat across the training boundary, indistinguishable from random
templates, and zero premise refutations outside the deepest-history
era. A weight-free structural generator finds engaged questions at
every cutoff, and adding the LLM back as a pure verbalizer refutes
premises in every era including the post-training one---locating the
foresight signal in pre-cutoff evidence structure, with the LLM as a
separable realization layer. We then validate the measurement
instrument itself with a seven-rater agreement study (two independent
blinded human annotators, five judge models, 90 items) and report what
it shows: two careful humans agree with each other at only
$\kappa=0.17$, every judge model agrees with the professional annotator
as well as or better than the humans agree with each other
($\kappa=0.17$--$0.26$), and frontier models agree with one another at
$\kappa=0.60$---so the common practice of certifying an LLM judge by
model--model agreement would have overstated its reliability threefold
here. The outcome taxonomy, not the judge, fails validation; absolute
rates are therefore rater-relative throughout, while the paper's
comparative claims are checked under three judges and survive with no
reversals. Two findings result: evidence-structure-first generation
outperforms LLM-only prompting at scale, and outcome taxonomies for
scientific-question evaluation need a measured human--human reliability
gate before any judge, human or model, is scored against them. A
prospective instance---200 questions from four generators, frozen
2026-08-17 with a 2027--2030 scoring window---is released so the
central claims become contamination-free tests that time itself will
grade.
\end{abstract}

\section{Introduction}
\label{sec:intro}

A growing family of systems claims to generate scientific research
questions, hypotheses, or ideas
\citep{lu2024aiscientist,wang2024scimon,baek2024researchagent}. How do we
know whether any of them is good at it? Today, essentially every
evaluation falls into one of three patterns: an \emph{expert score} (a
panel rates novelty and significance on a Likert scale), an \emph{LLM
score} (a language model rates the same properties), or a \emph{case
study} (a handful of generated ideas is narrated persuasively). All
three share the same defect: they are subjective. Expert panels disagree
with each other and with themselves \citep{si2024llmideas}; LLM judges
inherit the biases of their training distribution and can be steered by
phrasing; case studies are selected by the authors. None of these
evaluations can be \emph{wrong} in a way that data could demonstrate.
We hold our own instrument to that standard too: Section~\ref{sec:judgeval}
subjects it to a seven-rater reliability study and reports the result,
which is not flattering, in full.

Science itself offers a harder criterion. Research questions are bets
about where inquiry should go next, and the scientific community
eventually settles those bets: it invests observing time, funds
follow-ups, writes papers that answer some questions, poses others
independently, and refutes the premises of a few. This suggests a
measurement standard---deliberately a proxy, not a definition:

\begin{quote}
\emph{Future scientific engagement provides an observable, falsifiable
proxy for one important dimension of a question's value.}
\end{quote}

We do not claim engagement \emph{defines} value: community attention
carries popularity bias (Section~\ref{sec:threats}), and a question can
be excellent yet ignored for want of an instrument. The claim is
narrower and stronger where it counts---engagement is the one dimension
of value that is observable from frozen public data, and therefore the
one on which systems can be compared without asking anyone's opinion.

The standard becomes an evaluation protocol the moment we rewind the
clock. Fix a historical cutoff. Give a system only the literature
available before the cutoff. Freeze the questions it generates. Then let
the literature published \emph{after} the cutoff---which the system
never saw---grade the bet: Was the question answered? Substantially
advanced? Independently posed by working scientists? Ignored? Was its
underlying premise confirmed, or refuted? We call this procedure
\textbf{historical backtesting}, by analogy with the evaluation of
trading strategies on held-out past data \citep{bailey2014backtest}, and
with recent forecasting benchmarks that score models against events
occurring after training \citep{zou2022autocast}.

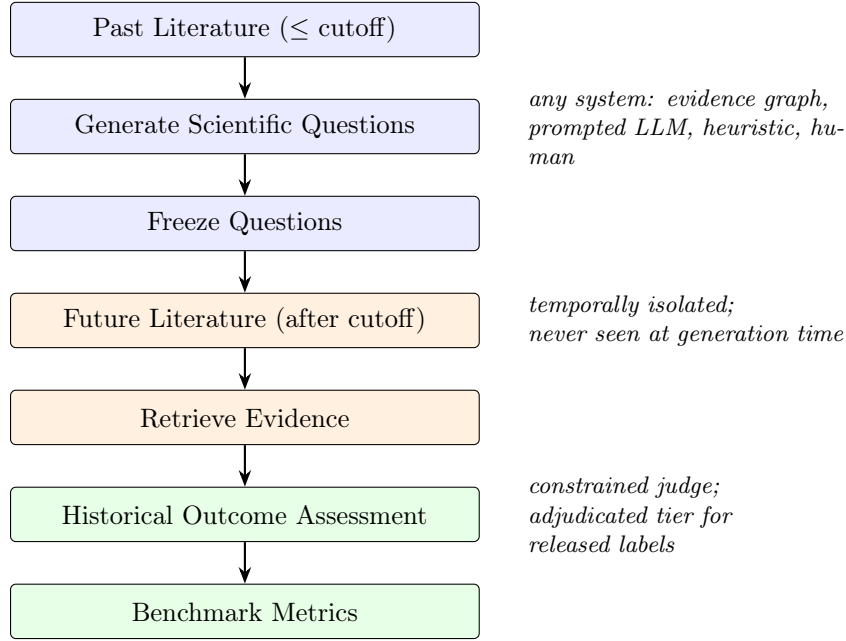
\begin{figure}[t]
\centering
\begin{tikzpicture}[
    node distance=0.55cm,
    stage/.style={draw, rounded corners=2pt, minimum width=6.2cm,
                  minimum height=0.72cm, align=center, font=\small},
    past/.style={stage, fill=blue!8},
    future/.style={stage, fill=orange!12},
    eval/.style={stage, fill=green!10},
    arr/.style={-{Stealth[length=2.2mm]}, thick}]
  \node[past]   (p1) {Past Literature ($\leq$ cutoff)};
  \node[past,   below=of p1] (p2) {Generate Scientific Questions};
  \node[past,   below=of p2] (p3) {Freeze Questions};
  \node[future, below=of p3] (f1) {Future Literature (after cutoff)};
  \node[future, below=of f1] (f2) {Retrieve Evidence};
  \node[eval,   below=of f2] (e1) {Historical Outcome Assessment};
  \node[eval,   below=of e1] (e2) {Benchmark Metrics};
  \draw[arr] (p1) -- (p2);
  \draw[arr] (p2) -- (p3);
  \draw[arr] (p3) -- (f1);
  \draw[arr] (f1) -- (f2);
  \draw[arr] (f2) -- (e1);
  \draw[arr] (e1) -- (e2);
  \node[right=0.5cm of p2, font=\footnotesize\itshape, text width=4.4cm, align=left]
    {any system: evidence graph,\\ prompted LLM, heuristic, human};
  \node[right=0.5cm of f1, font=\footnotesize\itshape, text width=4.4cm, align=left]
    {temporally isolated;\\ never seen at generation time};
  \node[right=0.5cm of e1, font=\footnotesize\itshape, text width=4.4cm, align=left]
    {constrained judge;\\ adjudicated tier for\\ released labels};
\end{tikzpicture}
\caption{The historical backtesting protocol. No question generator,
evidence graph, or particular LLM appears in the loop: the protocol
evaluates \emph{frozen question lists}, whatever produced them.}
\label{fig:protocol}
\end{figure}

Crucially, the protocol contains no question generator
(Figure~\ref{fig:protocol}). It takes a frozen list of questions as
input and returns outcome labels and metrics as output. Any
system---an evidence-graph pipeline, a prompted LLM, a citation
heuristic, a human scientist---can be evaluated under identical
conditions. This is what makes it a benchmark rather than a validation
appendix for one particular architecture.

\paragraph{Contributions.} We make five contributions; the first two
are claims ordered by strength, the remaining three are measurements
and artifacts:

\begin{enumerate}
  \item \textbf{Protocol (strong).} We formalize historical backtesting
    as an evaluation protocol for scientific question discovery:
    temporal isolation rules, a question-freezing requirement, fixed
    future-evidence retrieval, a two-dimensional outcome taxonomy that
    separates a question's fate (\labelfmt{answered},
    \labelfmt{partially\_addressed}, \labelfmt{posed\_but\_open},
    \labelfmt{not\_addressed}) from its premise's fate
    (\labelfmt{supported}, \labelfmt{refuted}, \labelfmt{weakened},
    \labelfmt{still\_plausible}, \labelfmt{not\_applicable}), and
    metrics defined over those labels (Sections~\ref{sec:protocol}
    and~\ref{sec:metrics}).
  \item \textbf{Benchmark instance (medium).} We release a reproducible
    astronomy instance with temporally isolated past and future corpora
    (2{,}512 and 1{,}891 papers; cutoff 2020-12-31), frozen questions,
    released retrieval records, adjudicated labels, one-command metric
    computation with CI-enforced reproducibility, a submission format,
    and four reference baselines evaluated through the identical
    pipeline (Sections~\ref{sec:dataset} and~\ref{sec:baselines}).
  \item \textbf{Empirical findings (cautious, and separated by sample
    size).} Our statistically supported claim concerns a \emph{class}
    of methods, not a single system: across 125-question submissions,
    evidence-structure-first generation resolves and refutes far more
    than direct LLM prompting (39\% vs.\ 15\% answered; 13\% vs.\ 0\%
    premise refutation, $p=3\times10^{-5}$), and this holds at four
    historical cutoffs including one whose future postdates the model's
    training data (Section~\ref{sec:foresight}). A weight-free
    structural generator---no LLM anywhere---outperforms LLM-only
    prompting on resolution, locating the foresight signal in
    pre-cutoff evidence structure rather than in model weights.
    Separately, and as an illustrative case rather than a statistical
    claim, a ten-question evidence-graph submission had every question
    engaged by later literature, one of them by refuting the premise it
    challenged (Section~\ref{sec:results}). At $n=10$ that submission
    cannot be ranked against baselines---detecting its apparent
    advantage would require $n\approx209$ per arm---and we make no such
    ranking claim. We do \emph{not} claim that AI reliably identifies
    the most valuable future scientific questions; we claim the
    protocol can tell us, eventually, whether it can, and that it
    already discriminates between generator families.
  \item \textbf{Measurement validity (adverse, and general).} We
    validate the measurement instrument itself with a seven-rater
    agreement study: two independent blinded human annotators and five
    judge models on 90 items. Humans agree with each other at
    $\kappa=0.17$; every model matches the professional annotator as
    well as the humans match each other; frontier models agree with one
    another at $\kappa=0.60$. The outcome taxonomy, not the judge,
    fails validation---and certifying an LLM judge by model--model
    agreement, the field's common shortcut, would have overstated
    reliability threefold here (Section~\ref{sec:judgeval}).
  \item \textbf{Prospective instance (frozen, unscoreable until 2031).}
    Two hundred questions from four generators, frozen at cutoff
    2026-08-17 with a pre-registered 2027--2030 scoring window and
    published corpus manifests and hashes---the contamination-free test
    that time itself will grade (Section~\ref{sec:conclusion}).
\end{enumerate}

What we believe is ultimately most useful here is not any single
result but the change of category: scientific question quality moves
from a matter of taste (``this question seems interesting'') to a
measured, comparable quantity (``under historical backtesting, 100\% of
this system's questions were engaged by later literature; 10\% led to a
premise refutation''). Every artifact needed to run the protocol---data,
code, labels, and checks---is public, and Astronomy~v1 is offered as the
first instance of the benchmark, not as its definition.

\section{Related Work}
\label{sec:related}

\paragraph{Automated scientific discovery and question generation.}
Computational discovery has a long lineage, from rule-based rediscovery
of physical laws \citep{langley1987} through closed-loop robot
scientists \citep{king2009} to the Nobel Turing Challenge's call for
AI scientists \citep{kitano2021}. Recent LLM-based systems generate
research ideas, hypotheses, or full papers: literature-based generation
\citep{wang2024scimon}, agentic idea refinement
\citep{baek2024researchagent}, and end-to-end automated research
\citep{lu2024aiscientist}. Literature-based discovery pioneered the
underlying intuition that recombining published evidence can anticipate
findings later verified empirically \citep{swanson1986}. Our work is
orthogonal to all of these: we do not propose a better generator; we
propose the missing evaluation.

\paragraph{Evaluating generated ideas.}
Existing evaluations are dominated by human preference and LLM scoring.
\citet{si2024llmideas} ran a large expert study comparing human and LLM
research ideas on rated novelty and excitement---the most rigorous
instance of the expert-score paradigm, and still a measurement of
\emph{opinion at generation time} rather than of what the ideas turned
out to be worth. LLM-as-judge scoring inherits known biases (position,
verbosity, self-preference) and, for questions about the future,
cannot be validated against ground truth at all. Historical backtesting
replaces both with an outcome variable that exists independently of any
rater: the subsequent behavior of the scientific community.

\paragraph{Backtesting and forecasting benchmarks.}
Scoring a strategy on held-out history is standard in quantitative
finance, along with well-documented failure modes---overfitting to the
backtest itself \citep{bailey2014backtest}---that motivate our
freezing and no-overwrite rules. Forecasting benchmarks score models on
events that resolve after training \citep{zou2022autocast}; retrodictive
evaluation with temporal holdouts is likewise used to test whether
models anticipate later discoveries \citep{swanson1986,king2009}.
We transplant this design to a harder target: not whether a stated
event occurs, but whether an open-ended research question earns the
community's future investment. Code benchmarks such as SWE-bench
\citep{jimenez2024swebench} demonstrated how a well-specified task
format plus frozen data can reorganize a research area around
measurable progress; we aim the same mechanism at question discovery.

\paragraph{Data contamination.}
Temporal splits are increasingly used to control LLM memorization in
evaluation. Our protocol controls the \emph{retrieval} channel
completely (corpus manifests, isolation rules, CI checks) and treats the
\emph{weights} channel---models whose training data postdates the
cutoff---as a declared, audited threat rather than a solved problem
(Section~\ref{sec:threats}).

\section{The Historical Backtesting Protocol}
\label{sec:protocol}

The protocol evaluates a set of frozen questions $Q = \{q_1,\dots,q_n\}$
against a future corpus. It has six steps; each is fully specified so
that two groups running the same instance obtain the same measurement.

\subsection{Step 1: Choose a cutoff}
A historical date $T$ (Astronomy~v1: 2020-12-31) splits the literature
into a \emph{past corpus} \corpusA{} (everything available up to $T$)
and a \emph{future window} realized as an isolated corpus \corpusB{}
(strictly after $T$). The cutoff must be far enough in the past for the
community to have had time to act---we recommend $\geq 4$ years---and
recent enough that the past corpus reflects a modern research frontier.

\subsection{Step 2: Generate questions}
Any method may generate questions: an evidence-graph pipeline, a
prompted LLM, a heuristic over citation statistics, a human expert. The
only requirements are (i) the generator consumes \emph{only} \corpusA{}
evidence, and (ii) every question records the pre-cutoff evidence it is
grounded in (\texttt{source\_evidence\_ids}). A submission whose source
evidence postdates $T$ is invalid, mechanically
(\texttt{scripts/validate\_cutoff.py}).

\subsection{Step 3: Freeze}
Questions are serialized---identifier, text, cutoff, generating system,
source evidence, system-assigned rank---with \texttt{frozen:\,true}
\emph{before any access to post-cutoff literature}, and are never edited
afterwards. Freezing is the protocol's load-bearing rule: without it,
question text drifts toward what the evaluator has meanwhile learned the
future contains, and the backtest silently becomes a description of the
future rather than a prediction of it \citep{bailey2014backtest}. In the
released implementation frozen files are append-only and guarded by CI;
editing a released question mints a new versioned instance rather than
overwriting the old one.

\subsection{Step 4: Define the future window}
\corpusB{} is collected under its own frozen manifest (query set, date
window, deduplication rules) and stored separately from \corpusA{};
records from \corpusB{} must never enter the generation pipeline. The
Astronomy~v1 window is 2021--2026. Bounding the window matters for
comparability: ``eventually engaged'' is not a fixed target, but
``engaged within $k$ years'' is.

\subsection{Step 5: Retrieve future evidence}
For each frozen question, the question text and every \corpusB{}
document (title + abstract) are embedded
(\texttt{text-embedding-3-small}); the top-$k$ documents by cosine
similarity ($k=8$) become the candidate evidence. Retrieval is
deliberately fixed and deliberately simple: systems are compared on
their \emph{questions}, not their retrievers, and reviewers can inspect
exactly which documents the judge saw because retrieval records are part
of the release.

\subsection{Step 6: Assess outcomes}
\label{sec:protocol:outcomes}

A judge reads the question and its retrieved candidates and assigns two
\emph{independent} labels (Table~\ref{tab:taxonomy}): the fate of the
question and the fate of its premise.

\begin{table}[t]
  \centering
  \small
  \caption{Outcome taxonomy v1.0. The two dimensions are labeled
  independently.}
  \label{tab:taxonomy}
  \begin{tabular}{@{}llp{8.2cm}@{}}
    \toprule
    Dimension & Label & Meaning \\
    \midrule
    \multirow{4}{*}{\texttt{outcome}}
      & \labelfmt{answered} & Future evidence substantially answers the question (including by refuting its premise) \\
      & \labelfmt{partially\_addressed} & Important, directly relevant progress; core question unresolved \\
      & \labelfmt{posed\_but\_open} & The community independently poses essentially the same question without resolving it \\
      & \labelfmt{not\_addressed} & No meaningful follow-up in the future corpus \\
    \midrule
    \multirow{5}{*}{\texttt{premise\_status}}
      & \labelfmt{supported} & Future evidence confirms the underlying premise \\
      & \labelfmt{refuted} & Future evidence falsifies the underlying premise \\
      & \labelfmt{weakened} & Substantial doubt cast without falsification \\
      & \labelfmt{still\_plausible} & The premise was not directly tested after the cutoff \\
      & \labelfmt{not\_applicable} & The question rests on no contestable premise \\
    \bottomrule
  \end{tabular}
\end{table}

Two design decisions deserve emphasis. First, the dimensions are
separated because the single most informative outcome a backtest can
surface---\emph{the community answered this question by refuting its
premise}---is inexpressible in a flat label set: it is simultaneously a
resolution (\labelfmt{answered}) and a falsification
(\labelfmt{refuted}). Our pilot's headline case
(Section~\ref{sec:results:case}) is exactly of this type. Second,
premise refutation is scored as a \emph{success} of the question, not a
failure: a question that provokes the community into overturning one of
its own published conclusions has done the most a question can do.

The judge operates under hard constraints enforced outside the model:
it may cite only bibcodes from the retrieved candidates (violations are
errors, never silently dropped); topical similarity is explicitly
insufficient---the cited paper must bear on the question's actual test
or premise; unknown labels fall back to the most conservative value and
are flagged. Labels then occupy one of two declared tiers.
\emph{Adjudicated} labels have passed human review under written
guidelines (citation validity, the engagement bar, the outcome/premise
split), with the adjudication log released; the headline labels of a
released instance are required to be of this tier (Astronomy~v1's are).
\emph{Judge-only} labels have not, are marked as such wherever
reported, and are the tier at which this paper's large-$n$ comparative
studies run (Sections~\ref{sec:scaled}--\ref{sec:foresight}). The
tier is part of every result's provenance; conflating them is a
protocol violation.

\subsection{Submissions}
A system is evaluated by submitting a directory containing
\texttt{questions.jsonl} (the frozen questions) and
\texttt{metadata.json} (system description, including a mandatory
declaration of any LLM components and their versions, for contamination
auditing). The benchmark pipeline---isolation checks, retrieval,
judging, adjudication where the tier requires it, metrics,
report---is identical for every
submission; the system whose questions we evaluate in
Section~\ref{sec:results} interacts with the benchmark only through
this interface.

\section{Benchmark Metrics}
\label{sec:metrics}

Let $Q$ be the $n$ frozen questions of a submission, with outcome labels
$o(q)$ and premise labels $p(q)$ as in Table~\ref{tab:taxonomy}, and let
$E(q)$ be the set of independent post-cutoff papers cited as supporting
evidence for $q$'s label. All rates are over $n$, so the four outcome
rates sum to one.

\begin{table}[t]
  \centering
  \small
  \caption{Benchmark metrics v1.0. All are computed by
  \texttt{benchmark/metrics.py} from the released annotation records.}
  \label{tab:metrics}
  \begin{tabular}{@{}lp{9.6cm}@{}}
    \toprule
    Metric & Definition \\
    \midrule
    Coverage (future attention rate) &
      $\frac{1}{n}\,|\{q : o(q) \neq \labelfmt{not\_addressed}\}|$ ---
      did future science engage the question at all? \\
    Answer rate &
      $\frac{1}{n}\,|\{q : o(q) = \labelfmt{answered}\}|$ \\
    Partial rate &
      $\frac{1}{n}\,|\{q : o(q) = \labelfmt{partially\_addressed}\}|$ \\
    Open rate &
      $\frac{1}{n}\,|\{q : o(q) = \labelfmt{posed\_but\_open}\}|$ ---
      the community independently recognized the question \\
    Premise refutation rate &
      $\frac{1}{n}\,|\{q : p(q) = \labelfmt{refuted}\}|$ ---
      questions that led to an established conclusion being overturned \\
    Evidence strength &
      mean $|E(q)|$ over engaged questions, and the multi-source rate
      $\frac{1}{n}\,|\{q : |E(q)| \geq 2\}|$ --- is the label supported
      by multiple independent papers? \\
    Lead time &
      years between question submission and the first time the community
      \emph{independently poses} the same question \\
    Community attention &
      volume of future investment engaging the question: papers,
      citations, review mentions, and major observing programs (e.g.\
      JWST/HST proposals) \\
    \bottomrule
  \end{tabular}
\end{table}

\paragraph{Coverage vs.\ answer rate.} Coverage asks whether the
question pointed anywhere the community went at all; the answer/partial/
open decomposition asks what happened when it got there. A system can
maximize coverage with fashionable-topic questions, which is why
coverage is never reported alone (Section~\ref{sec:threats} discusses
the popularity confound).

\paragraph{Premise refutation rate.} This is the metric we most want the
field to adopt. Questions that trigger refutations are the rarest and
arguably most valuable output of question discovery---they mark places
where the literature's accepted conclusions were wrong and where a
well-aimed question preceded the correction. Under the two-dimensional
taxonomy the refutation is recorded without erasing the fact that the
question was thereby \emph{answered}.

\paragraph{Lead time.} If a system poses a question at the cutoff and
the community first independently poses it in year $T{+}\ell$, the
system led the field by $\ell$ years; averaged over questions this
yields a comparable earliness score. Measuring $\ell$ requires
identifying \emph{community first-posed dates}, which demands careful
review-literature annotation we do not yet have. Astronomy~v1 therefore
reports \texttt{mean\_lead\_time\_years:\,null} rather than a number we
cannot defend; the released records do include a weaker, well-defined
lower bound---\emph{first-engagement lag}, the years from cutoff to the
earliest judge-cited supporting paper (pilot mean 2.9, range 1--5)---
which should not be confused with lead time.

\paragraph{Community attention.} Beyond binary engagement, the volume of
future investment (paper counts, citations to engaging papers, review
mentions, dedicated observing programs) reflects how much the community
cared. v1 records the ingredients (supporting bibcodes, their venues and
years) and reports evidence strength; a calibrated attention index is
future work, and Section~\ref{sec:threats} explains why raw attention
must never be the headline metric.

\paragraph{Reporting requirements.} A benchmark report must state: the
instance and protocol versions, $n$, all outcome and premise rates, the
evidence-strength pair, and either lead time or an explicit null. The
released implementation produces exactly this
(\texttt{results/astronomy\_v1/metrics.json}) with one command, and CI
fails if the committed numbers do not reproduce from the raw
annotations.

\section{The Astronomy v1 Instance}
\label{sec:dataset}

Astronomy~v1 instantiates the protocol in exoplanet atmospheres---a
domain chosen because it uniquely combines a fast-moving literature,
structured catalogs, and space-telescope archives, and because the
2021--2026 window contains a natural experiment: JWST began delivering
data mid-window, resolving questions that were unanswerable at the
cutoff. Table~\ref{tab:corpus} summarizes the instance.

\begin{table}[t]
  \centering
  \small
  \caption{Astronomy v1 at a glance. Corpus manifests, frozen questions,
  retrieval records, and adjudicated labels are all released.}
  \label{tab:corpus}
  \begin{tabular}{@{}ll@{}}
    \toprule
    Historical cutoff & 2020-12-31 \\
    Domain scope & atmospheric composition in transmission spectra;
                   cloud/haze degeneracies; instrument systematics \\
    \midrule
    \corpusA{} (past) & 2{,}512 deduplicated papers, 2015--2020 (NASA ADS;
                        500-paper full-text core) \\
    \corpusB{} (future) & 1{,}891 unique papers, 2021--2026, temporally
                          isolated \\
    Questions & 10, frozen, ranked, with pre-cutoff source evidence \\
    Retrieval & \texttt{text-embedding-3-small}, cosine, top-8; records released \\
    Judge & \texttt{gpt-4.1}, temperature 0, citation-constrained;
            human-adjudicated \\
    \bottomrule
  \end{tabular}
\end{table}

\paragraph{Corpora.} Both corpora are defined by frozen manifests---ADS
query sets, date windows, deduplication and filtering rules
(records without abstracts are dropped)---rather than by bulk data
dumps: the manifests are committed, and a script rebuilds either corpus
from its manifest via the ADS API. \corpusB{}'s manifest adds targeted
follow-up queries for the questions' objects (HD~189733, HD~209458,
WASP-12, WASP-121, TRAPPIST-1) so that engagement is measured against
the relevant future literature rather than against whatever a generic
query happens to return.

\paragraph{Leakage controls.} Temporal isolation is enforced
mechanically, not editorially: (1) nothing dated after the cutoff may
enter \corpusA{}, including catalog rows updated post-cutoff; (2) every
question's source evidence must predate the cutoff; (3) every retrieved
document must postdate it; (4) the judge may cite only retrieved
candidates; (5) question/retrieval/annotation records must align
one-to-one; (6) \corpusB{}'s window must start strictly after the
cutoff. All six checks run in continuous integration on every change to
the released data, together with a check that the released
\texttt{metrics.json} reproduces bit-identically from the raw
annotations.

\paragraph{Questions.} The ten frozen questions were generated by an
evidence-graph system \citep{paper1} from \corpusA{} only: claims with
provenance are extracted from the full-text core, cross-paper tensions
are detected and typed (observational tensions, methodological
challenges, single-dataset conclusions, independent qualifications), and
surviving signals are refined into ranked, falsifiable questions. For
the benchmark, that system is submission
\texttt{evidence\_graph\_v1}---evaluated through the same interface as
any future submission. Each released record carries the question text,
system rank, signal type, target objects, and pre-cutoff source
bibcodes; a curation log documenting human edits made \emph{before}
freezing (including a near-duplicate merge and presupposition fixes,
Section~\ref{sec:errors}) is released for audit.

\paragraph{What is released.} Frozen questions; corpus manifests; per-
question retrieval records (model, window, top-$k$, judge-cited
documents, top-1 similarity; full ranked lists with scores are scheduled
for v1.1); adjudicated outcome annotations with rationales and
supporting bibcodes; the adjudication log; computed metrics and
per-question results; four frozen baseline submissions with their full
retrieval records (per-document scores included), judge annotations,
and reports (Section~\ref{sec:baselines}); the scaled v1L instance
(manifests, configs, 424 frozen baseline questions, retrieval records
with per-document scores, judge annotations, per-system reports, and
the statistical comparison script of Section~\ref{sec:scaled}); the
tension-pair generators, the four-cutoff stress-test corpora manifests,
all 798 stress-test judgments, and the deterministic specificity rubric
of Section~\ref{sec:foresight}; and the complete pipeline code with
tests. The
data format is deliberately plain (JSONL + JSON manifests) so that other
groups can mint new instances---different domain, different cutoff---by
writing two manifests and one config file.

\section{Baselines}
\label{sec:baselines}

A benchmark that only ever scored one system would be a validation
appendix. Astronomy~v1 ships four reference baselines, chosen to
bracket the interesting comparisons; each consumes only \corpusA{}
records, freezes its output before any future-corpus access, and is
evaluated through the identical pipeline.

\begin{description}
  \item[B1: Random claims.] Sample random pre-cutoff papers and template
    their headline result into a robustness question. The floor: any
    system must beat chance-directed attention.
  \item[B2: Direct LLM.] Give an LLM (\texttt{gpt-4.1}, 60 sampled
    pre-cutoff abstracts) a request for the most valuable open
    questions. The ``why not just ask GPT?'' comparison. Because a
    modern LLM's weights postdate the cutoff, this baseline is also a
    contamination probe: performance that vanishes for
    post-training-cutoff instances indicates memorized hindsight rather
    than generation ability (Section~\ref{sec:threats}).
  \item[B3: Review future work.] Extract explicitly posed open questions
    from pre-cutoff review papers. The strongest natural reference: a
    discovery system is interesting only if it adds value over questions
    the community had already written down. Note this baseline should
    score highly on coverage \emph{by construction}---these questions
    are known community priorities---so the discriminating metrics are
    premise refutation and lead time, where a copied question can never
    lead the field.
  \item[B4: Citation leaders.] Template follow-up questions from the
    most-cited pre-cutoff papers. Tests whether chasing prominence
    matches structured evidence analysis. One caveat is built in: ADS
    citation counts are fetched at corpus-rebuild time and therefore
    include post-cutoff citations, so this baseline selects papers with
    \emph{hindsight} knowledge of which pre-2021 work the future found
    important---a bias in its favor that a cutoff-dated citation
    snapshot would remove.
\end{description}

\paragraph{Evaluation conditions.} All four baselines were run through
the released pipeline: top-8 retrieval with
\texttt{text-embedding-3-small} against the manifest-rebuilt future
corpus restricted to the frozen window (2{,}384 records, a superset of
the frozen 1{,}891 due to retroactive ADS indexing, containing all 22
citations in the released annotations), then \texttt{gpt-4.1} judging
at temperature 0 under the citation constraints of
Section~\ref{sec:protocol}. Baseline labels are \emph{judge-only}: they
have not received the human adjudication that the released
\texttt{evidence\_graph\_v1} labels did. To make that comparison
honest, Table~\ref{tab:leaderboard} also reports a judge-only rerun of
the evidence-graph submission under exactly the baseline conditions.
The rerun doubles as a replication check: it reproduces the released
mean top-1 retrieval similarity (0.668 vs.\ 0.666) and lands within one
label of the adjudicated results (coverage 90\% vs.\ 100\%, answered
10\% vs.\ 20\%, refutation 10\% = 10\%)---the deltas are exactly the
two labels adjudication had strengthened, so judge-only scoring reads
as the conservative floor of the adjudicated score.

\begin{table}[t]
  \centering
  \small
  \caption{Astronomy v1 leaderboard ($n=10$ questions per system).
  Adjud.\ = human-adjudicated labels; judge-only rows are directly
  comparable to each other. Cov.\ = future-attention rate; Ans.\ =
  answered; Part.\ = partially addressed; Open = posed but open; Ref.\ =
  premise refuted; $s_1$ = mean top-1 retrieval similarity. Lead time
  is null for all systems until community first-posed dates are
  annotated (Section~\ref{sec:metrics}).}
  \label{tab:leaderboard}
  \begin{tabular}{@{}llcccccc@{}}
    \toprule
    System & Eval & Cov. & Ans. & Part. & Open & Ref. & $s_1$ \\
    \midrule
    \texttt{evidence\_graph\_v1} \citep{paper1} & adjud. & 100\% & 20\% & 70\% & 10\% & 10\% & 0.666 \\
    \midrule
    \texttt{evidence\_graph\_v1} (rerun) & judge & 90\% & 10\% & 80\% & 0\% & 10\% & 0.668 \\
    B4 citation leaders     & judge & 100\% & 30\% & 70\% & 0\% & 0\% & 0.632 \\
    B3 review future work   & judge & 100\% & 10\% & 70\% & 20\% & 0\% & 0.593 \\
    B2 direct LLM           & judge & 90\% & 0\% & 90\% & 0\% & 0\% & 0.732 \\
    B1 random claims        & judge & 70\% & 0\% & 70\% & 0\% & 0\% & 0.611 \\
    \bottomrule
  \end{tabular}
\end{table}

\paragraph{Reading the leaderboard.} Three observations, offered with
the $n=10$ caution of Section~\ref{sec:results} applying to every row.

\emph{Coverage saturates.} Every non-random system scores 90--100\% on
future attention: in a field this active, any fluent, topical question
attracts partial engagement within five years. Coverage separates the
floor (random claims, 70\%, the only system with three
\labelfmt{not\_addressed} labels) from everything else, and nothing
else---which is why the protocol never reports it alone
(Section~\ref{sec:threats}).

\emph{Answered-rate comparisons need reading, not just ranking.} The
citation-leader baseline posts the highest judge-only answered rate
(30\%). Two mechanisms inflate it: its hindsight-biased paper selection
(above), and its template---``does the conclusion of highly cited paper
$X$ hold?''---which pattern-matches the replication studies that
prominent results reliably attract, so the judge can mark it answered
whenever the community re-examined a famous result for any reason. What
the template cannot do is risk anything: B4 refuted no premise, and by
construction a question of the form ``is the famous result right?''
poses nothing the community was not already testing. The evidence-graph
submission's answered questions, by contrast, specified particular
tests (Section~\ref{sec:results:case}) and include the leaderboard's
only refuted premise---on both adjudicated and judge-only rows.

\emph{The contamination probe registers a signal.} The direct-LLM
baseline has by far the highest retrieval similarity to future
literature ($s_1 = 0.732$ vs.\ 0.593--0.668 for every other system) and
the highest mean support count (4.6 cited papers per question)---its
questions are phrased in the way the 2021--2026 literature would come
to phrase them, consistent with weights that have read that literature.
Yet it resolves nothing: 0\% answered, 0\% refuted, 90\% partial. The
pattern suggests phrasing-level contamination without commitment to
falsifiable specifics---broad, well-aimed questions that everything
engages and nothing settles. Prospective instances
(Section~\ref{sec:conclusion}) will separate the two channels
definitively.

Baseline generators, frozen submissions, retrieval records with full
per-document scores, judge annotations, and per-system reports are all
released; the leaderboard file is designed for external submissions to
append to. All three observations above are drawn from ten questions
per system; Section~\ref{sec:scaled} re-examines them on a scaled
instance with 424 baseline questions, and two of the three require
revision there.

\section{Pilot Results: Historical Validation}
\label{sec:results}

We ran the full protocol on the ten frozen
\texttt{evidence\_graph\_v1} questions. Headline numbers: every
question was substantively engaged by the 2021--2026 literature
(coverage 100\%); two were answered, seven partially addressed, one
independently posed and still open; one premise was refuted. Evidence
strength: 2.4 supporting papers per question on average, with 60\% of
labels supported by $\geq 2$ independent papers. The earliest
judge-cited engagement came 1--5 years after the cutoff (mean 2.9).
Table~\ref{tab:perq} gives the per-question picture.

\begin{table}[t]
  \centering
  \small
  \caption{Per-question outcomes for \texttt{evidence\_graph\_v1} on
  Astronomy v1 (system rank order; $|E|$ = independent supporting
  papers; year = earliest cited engagement).}
  \label{tab:perq}
  \begin{tabular}{@{}clllcc@{}}
    \toprule
    Rank & ID & Outcome & Premise & $|E|$ & Year \\
    \midrule
    1  & q\_002 & \labelfmt{posed\_but\_open}      & \labelfmt{still\_plausible} & 1 & 2022 \\
    2  & q\_001 & \labelfmt{partially\_addressed}  & \labelfmt{supported}        & 3 & 2021 \\
    3  & q\_004 & \labelfmt{partially\_addressed}  & \labelfmt{supported}        & 1 & 2021 \\
    4  & q\_008 & \labelfmt{answered}              & \labelfmt{refuted}          & 3 & 2025 \\
    5  & q\_007 & \labelfmt{partially\_addressed}  & \labelfmt{still\_plausible} & 3 & 2024 \\
    6  & q\_010 & \labelfmt{partially\_addressed}  & \labelfmt{still\_plausible} & 1 & 2022 \\
    7  & q\_005 & \labelfmt{answered}              & \labelfmt{supported}        & 2 & 2024 \\
    8  & q\_009 & \labelfmt{partially\_addressed}  & \labelfmt{supported}        & 5 & 2021 \\
    9  & q\_006 & \labelfmt{partially\_addressed}  & \labelfmt{still\_plausible} & 1 & 2025 \\
    10 & q\_011 & \labelfmt{partially\_addressed}  & \labelfmt{still\_plausible} & 4 & 2024 \\
    \bottomrule
  \end{tabular}
\end{table}

\subsection{Case study: a premise refuted (q\_008, HD 209458 b)}
\label{sec:results:case}

From pre-2021 evidence, the system flagged a single-dataset conclusion:
the influential retrieval of a \emph{strongly subsolar} water abundance
at the terminator of HD~209458\,b \citep{macdonald2017}. The frozen
question asked, in 2020 terms:

\begin{quote}\itshape
Is the strongly subsolar terminator water abundance retrieved for
HD~209458\,b a property of the atmosphere or an artifact of retrieval
assumptions, as tested by comparing independent retrieval frameworks on
the same and on independent datasets?
\end{quote}

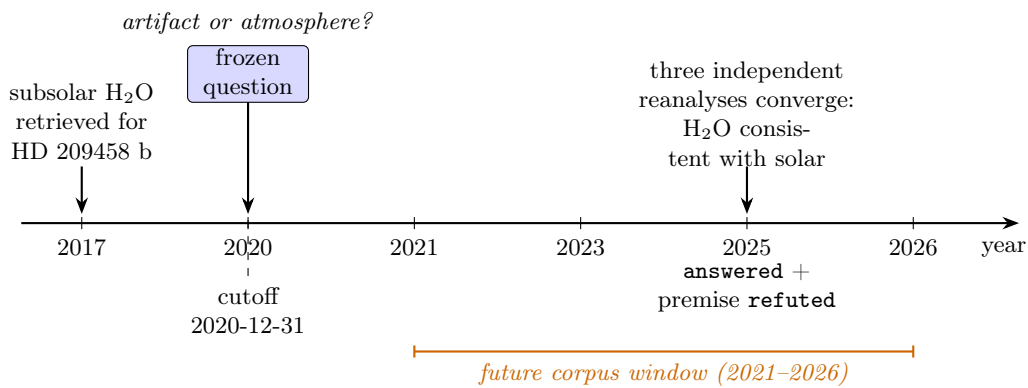
\begin{figure}[t]
\centering
\begin{tikzpicture}[
    font=\small,
    evt/.style={align=center, font=\footnotesize},
    arr/.style={-{Stealth[length=2.2mm]}, thick}]

  \draw[arr] (0,0) -- (13.2,0);
  \node[font=\footnotesize] at (13.0,-0.35) {year};
  \foreach \x/\y in {0.8/2017, 3.0/2020, 5.2/2021, 7.4/2023, 9.6/2025, 11.8/2026}
    \draw (\x,0.08) -- (\x,-0.08) node[below, font=\footnotesize] {\y};

  \node[evt, text width=2.6cm] at (0.8,1.35)
    {subsolar H$_2$O\\ retrieved for\\ HD 209458 b};
  \draw[arr] (0.8,0.75) -- (0.8,0.12);

  \draw[fill=blue!15, rounded corners=2pt] (2.2,1.6) rectangle (3.8,2.35);
  \node[evt] at (3.0,1.975) {frozen\\ question};
  \draw[arr] (3.0,1.6) -- (3.0,0.12);
  \node[evt] at (3.0,2.65) {\emph{artifact or atmosphere?}};

  \draw[dashed] (3.0,-0.7) -- (3.0,1.4);
  \node[evt] at (3.0,-1.15) {cutoff\\ 2020-12-31};

  \node[evt, text width=3.4cm] at (9.6,1.45)
    {three independent\\ reanalyses converge:\\ H$_2$O consistent with solar};
  \draw[arr] (9.6,0.75) -- (9.6,0.12);
  \node[evt, text width=3.9cm] at (9.6,-0.85)
    {\labelfmt{answered} + premise \labelfmt{refuted}};

  \draw[thick, orange!80!black] (5.2,-1.7) -- (11.8,-1.7)
    node[midway, below, font=\footnotesize\itshape]
    {future corpus window (2021--2026)};
  \draw[thick, orange!80!black] (5.2,-1.78) -- (5.2,-1.62);
  \draw[thick, orange!80!black] (11.8,-1.78) -- (11.8,-1.62);
\end{tikzpicture}
\caption{Timeline of q\_008. The question was frozen from pre-2021
evidence; by 2025 three independent analyses had performed the test it
specified and refuted its premise---the strongly subsolar water
abundance was a retrieval artifact.}
\label{fig:case}
\end{figure}

The 2021--2026 literature then performed exactly the test the question
specified (Figure~\ref{fig:case}): a reanalysis of HST and JWST spectra
with improved systematics treatment and Bayesian model averaging, an
independent retrieval framework demonstrating that free-vs-equilibrium
chemistry assumptions span the subsolar-to-solar range, and ground-based
high-resolution spectroscopy constraining the abundance independently of
space-based data. All three converge: the terminator water abundance is
consistent with solar, and the strongly subsolar value was an artifact
of earlier retrieval assumptions and data systematics. Under the
taxonomy this is \labelfmt{answered} + \labelfmt{refuted}: the question
was resolved \emph{by} the community overturning the premise the
question challenged---an outcome that no expert score assigned in 2020
could have certified, and precisely what backtesting exists to detect.

\subsection{Secondary observations}

\paragraph{The top-ranked question is independently posed and open.}
The system's rank-1 question (q\_002: how terminator heterogeneity
biases the WASP-12\,b water abundance and C/O ratio) tracks a
methodological concern the community has since engaged in general form
--- inhomogeneous-terminator biases in retrievals --- without resolving
it for WASP-12\,b specifically: \labelfmt{posed\_but\_open}. A question
the field poses but has not answered is a live research target; that the
system's top pick lands there is the behavior a ranking is supposed to
produce, though $n=1$ at rank 1 proves nothing by itself.

\paragraph{An answered null result.} q\_005 asked whether HST
transmission data independently support NH$_3$ or HCN in HD~209458\,b
--- a molecular-detection claim from the same single-dataset analysis as
q\_008 \citep{macdonald2017}. By 2024--2025, high-resolution
spectroscopy had placed stringent upper limits on both species:
\labelfmt{answered}, premise \labelfmt{supported} (the data indeed do
not independently support the detection). Backtesting counts a
cleanly resolved null exactly as it counts a positive.

\paragraph{Instrument-gated engagement.} q\_011 (whether JWST validated
pre-launch predictions of TRAPPIST-1 CO$_2$ detectability,
\citealp{lustigyaeger2019}) could not have been engaged before JWST flew;
its first cited engagement is 2024 and stellar contamination has so far
prevented a definitive test. Engagement timing is partly an instrument
schedule, not purely a question-quality signal---a confound
Section~\ref{sec:threats} treats explicitly.

\paragraph{What these ten questions do not show.} With ten questions
from one system in one domain, rates carry wide intervals (the exact
95\% Clopper--Pearson interval for 10/10 coverage is $[0.69, 1.0]$),
the baseline rows are judge-only rather than adjudicated
(Section~\ref{sec:baselines}), and the generating system's LLM
components postdate the cutoff (Section~\ref{sec:threats}). Moreover,
Section~\ref{sec:judgeval} shows all absolute rates are
rater-relative: ``10/10 engaged'' is this instance's adjudicated
reading, made by the authors, not a rater-free fact. The pilot
demonstrates that the protocol runs end-to-end, yields auditable,
reproducible labels, and prices its baselines; it does not establish
that any system reliably anticipates future science.

\section{Scaling the Baselines: Astronomy v1L}
\label{sec:scaled}

Every baseline conclusion in Section~\ref{sec:baselines} rests on ten
questions per system. To test which of them survive a larger sample, we
minted a second instance, \textbf{astronomy v1L}: same cutoff, same
retrieval and judge settings, but corpora rebuilt from broadened
manifests (12 past-corpus and 18 future-corpus ADS query sets covering
clouds and hazes, atmospheric escape, phase curves, high-resolution
spectroscopy, and JWST; 4{,}040 past and 5{,}754 frozen-window future
records---1.8$\times$ and 2.4$\times$ the v1 corpora) and baselines
scaled to 125 questions each. The review--future-work extractor is the
exception by necessity: it exhausts the supply of explicitly posed
questions in all 4{,}040 pre-cutoff abstracts at 49---itself a finding;
the community's already-written-down questions are a finite resource.
In total v1L evaluates 424 frozen baseline questions plus the ten
evidence-graph questions re-run as a cross-instance anchor, all
judge-only. The v1 instance and its released data are untouched.

\begin{table}[t]
  \centering
  \small
  \caption{Astronomy v1L results (judge-only). Brackets are exact 95\%
  Clopper--Pearson intervals. Eng.\ = future-attention rate; Ans.\ =
  answered; Ref.\ = premise refuted; $s_1$ = mean top-1 retrieval
  similarity; lag = mean years to earliest cited engagement. The
  evidence-graph row is the ten v1 questions re-evaluated on the v1L
  corpus (anchor), not a scaled submission.}
  \label{tab:scaled}
  \begin{tabular}{@{}lrllllr@{}}
    \toprule
    System & $n$ & Eng. & Ans. & Ref. & $s_1$ & lag \\
    \midrule
    \texttt{evidence\_graph} (anchor) & 10 & 80\% [44,97] & 30\% [7,65] & 10\% [0,45] & 0.682 & 3.5 \\
    B2 direct LLM & 125 & 96\% [91,99] & 15\% [9,23] & 0\% [0,3] & 0.722 & 1.9 \\
    B4 citation leaders & 125 & 87\% [80,93] & 26\% [18,34] & 3.2\% [1,8] & 0.632 & 2.5 \\
    B3 review future work & 49 & 92\% [80,98] & 22\% [12,37] & 0\% [0,7] & 0.567 & 2.4 \\
    B1 random claims & 125 & 73\% [64,80] & 11\% [6,18] & 0\% [0,3] & 0.622 & 2.8 \\
    \bottomrule
  \end{tabular}
\end{table}

Table~\ref{tab:scaled} gives the scaled results. Sample size changes
two of Section~\ref{sec:baselines}'s three conclusions and sharpens the
third---which is the point of running the experiment.

\paragraph{Revised: coverage does discriminate at scale.} At $n=10$
every non-random system sat at 90--100\% engagement and we concluded
coverage separates only the floor. At $n=125$ the rates pull apart:
random claims 72.8\% [64,80], citation leaders 87.2\%, direct LLM
96.0\% [91,99]; random vs.\ direct-LLM engagement differs at
$p<10^{-4}$ and random vs.\ citation leaders at $p=0.007$ (two-sided
Fisher). The v1 reading was a small-sample artifact. What survives is
the ceiling: fluent, topical LLM questions still approach saturation,
so coverage separates the bottom and middle of the range while
compressing the top---it remains unusable as a sole metric.

\paragraph{Revised: premise refutation is not unique to the
evidence-graph system.} At $n=10$ no baseline refuted a premise; at
$n=125$ the citation-leader baseline catches four refutations (3.2\%
[1,8]): the subsolar-water conclusion for HD~209458\,b (the same
\citealp{macdonald2017} result behind q\_008), the methane-depleted
atmosphere of K2-18\,b overturned by JWST, TiO in WASP-121\,b
unconfirmed by later data, and systematic bias found in benchmark
ultracool-dwarf retrievals. Asking ``is the famous result right?''\ of
enough famous results does eventually catch the ones that fall---note
its hindsight-biased selection (Section~\ref{sec:baselines}) works in
its favor here, since post-cutoff citation counts are inflated by
exactly the controversies that produce reversals. Three things remain
true. Refutation is the rarest outcome for every system (random
templates: 0/125, upper bound 2.9\%---refutations are not free; a
question must aim at a contestable claim). The evidence-graph
submission's nominal rate stays highest (10\% vs.\ 3.2\%), but $n=10$
cannot establish superiority ($p=0.32$; the comparison needs
$n\approx209$ per arm for 80\% power, Section~\ref{sec:threats}). And the two routes to a
refutation differ qualitatively: prominence-chasing rediscovers that
famous claims attract scrutiny, while the evidence-graph question
specified the decisive test from pre-cutoff evidence tensions
(Section~\ref{sec:results:case}). The scaled data cannot yet separate
those routes quantitatively; a scaled evidence-graph submission could.

\paragraph{Sharpened: there is a nonzero answered floor.} Random
robustness templates get \emph{answered} 11.2\% [6,18] of the
time---the field re-examines even arbitrarily chosen results at a
measurable base rate. The v1 estimate of that floor (0/10) was too
flattering to every other system: an answered rate is meaningful only
against $\sim$11\%, not zero. Citation leaders (25.6\%) clear the floor
($p=0.005$); direct LLM (15.2\%) does not ($p=0.42$).

\paragraph{Persistent: the contamination signature.} The direct-LLM
baseline keeps the highest similarity to future literature at scale
($s_1=0.722$ vs.\ 0.567--0.682 for all others), the broadest engagement
(96\%, multi-source rate 94\%), and the earliest mean engagement (1.9
years---its cited evidence concentrates in 2021--2022, the years
closest to its training distribution). Scaling revises one part of the
v1 reading: it does convert engagement into answers (15.2\% vs.\ 0/10
at $n=10$), but at a rate statistically indistinguishable from random
templates, despite engaging twice as much of the literature. Breadth
without resolution remains the signature.

\paragraph{Cross-instance anchor: metrics are corpus-relative.}
Re-judging the ten v1 questions on the 2.4$\times$ larger corpus flips
four labels in both directions: two questions gain \labelfmt{answered}
(richer evidence pools), one drops to \labelfmt{not\_addressed} (its
engaging papers pushed out of a top-8 that a larger corpus makes more
competitive), and one premise moves \labelfmt{weakened}$\to$
\labelfmt{still\_plausible}. q\_008's \labelfmt{answered} +
\labelfmt{refuted} reproduces. The lesson is structural: rates are
functions of the (corpus, retriever, judge) triple, so rows are
comparable only within an instance---v1 and v1L rows must never be
ranked against each other, and the frozen-instance design exists
precisely to make the triple explicit.

\paragraph{What scaling did and did not change.} The scaled study
strengthens the benchmark's discriminative claims (coverage now
separates three tiers; answered rates have a measurable floor) and
weakens one system-level claim (refutation exclusivity). It does not
change the evidence-graph submission's standing---its rates are
unchanged and its refutation reproduces---but it narrows what that
standing demonstrates: at current sample sizes, the defensible
statement is that structured evidence analysis found a refutation by
specifying its test in advance, not that it finds refutations at a
higher rate than strong heuristics. Settling the rate question requires
scaling the \emph{submission}, not just the baselines---the first item
on the revised roadmap.

\section{Separating Hindsight Memorization from Foresight}
\label{sec:foresight}

The deepest objection to any backtest run with a modern LLM anywhere in
the loop (Section~\ref{sec:threats}) is that its apparent performance
decomposes into three terms:
\begin{equation*}
\text{Performance} \;=\;
\underbrace{\text{reasoning over pre-cutoff evidence}}_{\text{what we want}}
\;+\; \underbrace{\text{memorized future}}_{\text{contamination}}
\;+\; \underbrace{\text{topic prior}}_{\text{fashion}}
\end{equation*}
and a single retrospective instance cannot tell the terms apart. A
question that 2024 answered may have been \emph{predicted} from 2020
evidence---or \emph{remembered} from the model's training data. This
section reports two experiments designed to pry the terms apart:
holding the cutoff fixed while varying which component generates the
question, and holding the generators fixed while moving the cutoff
across the judge-model's training boundary.

\subsection{Same cutoff, different generators: where does the signal
live?}
\label{sec:foresight:abc}

Three generation pipelines share the 2020 cutoff and the v1L evaluation
conditions but differ in what produces the question:

\begin{description}
  \item[A --- LLM only.] The direct-LLM baseline: \texttt{gpt-4.1} reads
    sampled pre-cutoff abstracts and proposes questions. Weights fully
    exposed to post-2020 literature.
  \item[B --- structure $\to$ LLM.] A deterministic, LLM-free detector
    finds pairs of pre-cutoff abstracts about the same catalogued object
    with opposing stances on the same species (detection vs.\
    non-detection / upper limit); \texttt{gpt-4.1}'s only job is to
    verbalize each detected tension as one falsifiable question. The
    evidence structure is fixed before any LLM sees anything.
  \item[C --- structure only.] The same detected pairs rendered by a
    fixed template. No LLM anywhere: this generator has no weights to
    contaminate.
\end{description}

B and C share identical evidence structures, so their gap isolates the
language-realization layer; A and B share the same LLM, so their gap
isolates evidence structure. (B/C are an evidence-structure-\emph{lite}
probe---object co-mention plus stance cues---not a reimplementation of
the evidence-graph system.) Table~\ref{tab:abc} gives the $n=125$
results.

\begin{table}[t]
  \centering
  \small
  \caption{The A/B/C decomposition on astronomy v1L (judge-only,
  $n=125$ each; brackets are 95\% Clopper--Pearson intervals).
  $s_{\mathrm{obj}}$ = share of questions naming a specific catalogued
  object (deterministic rubric, Section~\ref{sec:foresight:saf}).}
  \label{tab:abc}
  \begin{tabular}{@{}lllllr@{}}
    \toprule
    Pipeline & Eng. & Ans. & Ref. & $s_1$ & $s_{\mathrm{obj}}$ \\
    \midrule
    A: LLM only              & 96\% [91,99] & 15\% [9,23]  & 0\% [0,3]   & 0.722 & 5\% \\
    B: structure $\to$ LLM   & 74\% [65,81] & 39\% [31,48] & 13\% [8,20] & 0.694 & 95\% \\
    C: structure only        & 77\% [68,84] & 25\% [17,33] & 2.4\% [0,7] & 0.721 & 100\% \\
    \bottomrule
  \end{tabular}
\end{table}

Three readings. \emph{First}, the contamination-free pipeline C beats
the fully exposed pipeline A on answered rate (24.8\% vs.\ 15.2\%,
$p=0.08$) and on refutations (3 vs.\ 0)---evidence that a foresight
signal exists in pre-cutoff evidence structure alone, extractable with
zero model weights. \emph{Second}, adding the LLM back as a pure
verbalizer (B) roughly doubles resolution over the same structure
(39.2\% vs.\ 24.8\% answered, $p=0.02$---suggestive only; this
contrast does not survive the multiple-comparison correction of
Section~\ref{sec:threats}---and 12.8\% vs.\ 2.4\% refuted, $p=0.003$,
which does): phrasing a tension as a crisp either/or question makes it
judgeable, and B ends with five times pipeline A's refutation count
while citing the same model. \emph{Third}, A's questions are
structurally different, not just weaker: only 5\% name a specific
object (vs.\ 95--100\% for B/C), and its engagement is the highest of
any system---broad questions that everything touches and little
settles.

\subsection{Same generators, moving cutoff: the temporal stress test}
\label{sec:foresight:stress}

If pipeline A's performance were substantially the memorized-future
term, it should degrade as the cutoff crosses the model's training
boundary (June 2024 for \texttt{gpt-4.1}). We ran four generators (A,
B, C, and random claims as an era control) at four cutoffs---2010,
2015, 2020, 2024---with uniform four-year future windows ($n=50$ per
cell; the 2024 window is censored at 2026-06 and flagged; corpora per
era rebuilt from released manifests). Table~\ref{tab:stress} reports
the grid.

\begin{table}[t]
  \centering
  \small
  \caption{Temporal contamination stress test (judge-only, $n=50$ per
  cell; c2010 tension cells have $n=48$---the 855-paper 2005--2010
  corpus yields only 48 detectable tension pairs). Cutoffs 2010--2020
  lie inside the LLM's training data; the 2024 cutoff's future window
  (2025--2026-06, censored) postdates it. Ans.\ = answered; Ref.\ =
  refuted count.}
  \label{tab:stress}
  \begin{tabular}{@{}llcccc@{}}
    \toprule
    & & c2010 & c2015 & c2020 & c2024 \\
    \midrule
    A: LLM only & Ans. & 22\% & 12\% & 10\% & 16\% \\
                & Ref. & 2 & 0 & 0 & 0 \\
    B: structure $\to$ LLM & Ans. & 31\% & 26\% & 34\% & 32\% \\
                & Ref. & 6 & 3 & 6 & 5 \\
    C: structure only & Ans. & 10\% & 6\% & 16\% & 4\% \\
                & Ref. & 0 & 0 & 0 & 0 \\
    Random claims & Ans. & 12\% & 6\% & 4\% & 6\% \\
                & Ref. & 1 & 1 & 0 & 0 \\
    \bottomrule
  \end{tabular}
\end{table}

\paragraph{The naive collapse does not happen.} Pipeline A's answered
rate is statistically flat across the training boundary (14.7\% pooled
in-training vs.\ 16.0\% post-training, $p=0.82$), as is every other
generator's. At the outcome level, the memorized-future term is
\emph{not} where A's performance comes from---because, as the
decomposition shows, A's performance never rested on specifics that
memorization could supply. Its engagement sits at 92--98\% at every
cutoff, its specificity at the floor ($\approx$1.1 of 3) at every
cutoff: broad questions about each era's active topics, engaged
everywhere, resolving little, in any era. That is the topic-prior term
at work, and a topic prior does not need to remember the future---the
present is enough.

\paragraph{Where a memorization trace does appear.} Two places, both
in the channels the topic prior cannot supply. Pipeline A's only
premise refutations in the entire stress test (2 of 200) occur at the
deepest-contamination cutoff, 2010---the one era whose reversals the
model has certainly read about---and never after (0 of 150; too few
for significance, CI [0.5\%, 13.7\%] at c2010). And A's phrasing
similarity to future literature is highest inside its training window
(0.717--0.724) with a mild post-training dip (0.703)---directionally
consistent with phrasing-level memorization, though era confounds keep
this suggestive rather than conclusive.

\paragraph{The structural signal is era-robust.} Pipeline B refutes
premises at every cutoff---6, 3, 6, 5---including the one whose future
the model cannot have seen (10.0\% post-training vs.\ 10.1\%
in-training). Across all cutoffs B refutes at 10.1\% vs.\ A's 1.0\%
($p=4\times10^{-5}$), and B's answered margin over C persists
post-training ($+28$ points at c2024). One residual channel remains
open and is worth stating precisely: c2024 tension pairs are drawn from
2019--2024 abstracts, and a tension resolved in early-2024 literature
the model saw could steer B's phrasing even though the evaluation
window postdates training. C is immune by construction, which is why C
finding \emph{any} future engagement at every cutoff (70--83\%) is the
cleanest single fact in the grid.

\subsection{Specificity-adjusted foresight}
\label{sec:foresight:saf}

Coverage can be farmed by asking broad questions---``How can we better
understand exoplanet atmospheres?'' will be engaged with probability 1
in any active field. We therefore score every question with a
deterministic, released rubric: $+1$ for naming a specific catalogued
object, $+1$ for naming a measurable claim (species, quantity with
units, abundance comparative), $+1$ for an explicit discriminative
construction (``\ldots or an artifact of\ldots'', ``as tested by'');
and define specificity-adjusted foresight
$\mathrm{SAF} = \mathbb{E}[\,w(\text{outcome}) \cdot \text{specificity}/3\,]$
with $w$ = 1 / 0.5 / 0.25 / 0 for answered / partial / posed-open /
not addressed. (Template pipelines inherit the test-construction
point from their template---the rubric's components are reported
separately for exactly that reason; a novelty term awaits first-posed
dates, Section~\ref{sec:conclusion}.) Two facts survive every era and
instance: pipeline A's anchoring is an order of magnitude below the
structure pipelines' ($s_{\mathrm{obj}}$ 5\% vs.\ 95--100\% on v1L),
and its SAF never exceeds the random-template floor by more than a few
points (0.18--0.23 vs.\ 0.24--0.26), while structure pipelines reach
0.34--0.43. The evidence-graph submission's ten questions score
$s_{\mathrm{obj}}=90\%$, SAF 0.40.

\subsection{What this section establishes}

\begin{quote}
\emph{Memorized relevance is not scientific foresight.}
\end{quote}

The LLM-only pipeline exhibits relevance everywhere---near-ceiling
engagement, the closest phrasing to the future literature at every
cutoff---and specific foresight nowhere: no refutation outside the era
it could have memorized, an answered rate indistinguishable from random
templates, specificity at the floor. The structure-first pipelines
invert the picture: lower engagement, higher resolution, refutations in
every era including the one that postdates the model's training. On
this evidence, the foresight signal measured by historical backtesting
lives chiefly in pre-cutoff evidence structure; the LLM contributes a
real but separable service---turning a detected tension into a
question sharp enough to be judged. Contamination, meanwhile, turns
out to be measurable rather than merely confessable: the stress test
bounds its outcome-level effect (small for this task family) and
localizes its traces (phrasing proximity; refutations only in deep
history). All 798 stress-test judgments, the A/B/C submissions, and
the per-cell corpora manifests are released; every number in this
section is judge-only and carries $n=48$--$125$ intervals---the
qualitative pattern, not any single rate, is the finding.

\section{Judge Validation}
\label{sec:judgeval}

Replacing expert scores with an LLM judge only helps if the judge is
itself accountable. This section reports four validation experiments on
a frozen 90-question sample, stratified across systems and outcome
labels (\texttt{results/judge\_validation/}), plus a seven-rater
agreement study built on the released blinded annotation apparatus.
The internal checks pass, or fail in ways the data explain; the
external one --- two independent human annotators against five judge
models --- does not, and it reframes what an LLM judge can even be
validated against (Section~\ref{sec:judgeval:human}).

\subsection{Does the judge discriminate, or merely detect topic?}
\label{sec:judgeval:mismatch}

The sharpest failure mode for an engagement metric is that it measures
topical similarity and calls it engagement. We test this directly:
re-judge every sampled question after swapping in \emph{another
question's} retrieved evidence. A discriminating judge must collapse to
\labelfmt{not\_addressed}.

Engagement falls from 74\% (67/90) on true pairings to 20\% (18/90) on
mismatched ones ($p<10^{-4}$, two-sided Fisher), and the drop is
individually significant for five of six systems. Two further readings
matter. The 20\% residual is a \emph{generosity bound}: on evidence
that cannot possibly bear on the question, this judge still reports
engagement one time in five, so every engagement rate in this paper
should be read against that floor rather than against zero. And the
drop tracks question specificity---cleanest for the most anchored
questions (tension-LLM 12/16 $\to$ 1/16, $p=0.0002$; evidence-graph
8/10 $\to$ 1/10) and weakest for the generic citation-leader template
(11/16 $\to$ 5/16, $p=0.076$, the only non-significant cell). A
question vague enough to accept unrelated evidence is vague enough to
fool the judge, which is independent support for the specificity rubric
of Section~\ref{sec:foresight:saf}.

\subsection{Judge stochasticity and prompt sensitivity}

Temperature 0 is not determinism. Re-running the identical
configuration twice gives outcome agreement 96.7\% and 93.3\%
($\kappa=0.95$, $0.91$) and premise agreement 95.6\% and 94.4\%
($\kappa=0.92$, $0.90$): roughly 3--7 points of label noise, small
relative to the effects the paper reports but not zero, and it should
be assumed present in every rate.

Prompt sensitivity is larger. A semantically equivalent rewrite of the
judge prompt agrees at $\kappa=0.68$ (outcome) and $0.72$ (premise). A
harder variant, replacing the label \emph{names} with neutral codes
(\texttt{L1}--\texttt{L4}, \texttt{P1}--\texttt{P5}) while keeping the
definitions verbatim, drops to $\kappa=0.57$ and $0.55$, and never once
uses the code corresponding to \labelfmt{posed\_but\_open}. Part of the
judge's behaviour therefore rests on the connotations of the label
names, not on their stated definitions. We report this as a real
limitation: label naming is part of the protocol and must be frozen
along with everything else.

\subsection{Cross-model agreement, and what low agreement means here}

Re-judging with \texttt{gpt-4o} and \texttt{gpt-4-turbo} gives low
nominal agreement with \texttt{gpt-4.1}: pairwise Cohen's
$\kappa=0.34$ and $0.20$ on outcome, $0.19$ and $0.05$ on premise
status; three-way Fleiss $\kappa=0.38$ and $0.03$. Taken alone these
numbers say the instrument is unreliable, and we report them
unadorned.

The label distributions complicate that reading. On 90 items
\texttt{gpt-4-turbo} assigns \labelfmt{still\_plausible} 87 times and
never once uses \labelfmt{refuted} or \labelfmt{weakened};
\texttt{gpt-4o} assigns it 78 times. On the outcome dimension both
almost never use \labelfmt{answered} (2 and 4 times of 90, against 25
for \texttt{gpt-4.1}), collapsing a four-way judgement into a
two-way one. Their comparatively high mutual agreement
($\kappa=0.71$ on outcome) is therefore consistent with a shared
conservative default rather than with shared judgement, and $\kappa$ is
in any case depressed when one rater's marginals are near-degenerate.

We flag plainly that this reading is self-serving---``the judges who
disagree with ours are the incompetent ones'' is exactly what a
motivated author would say---and that only human annotation can
arbitrate it. Section~\ref{sec:judgeval:human} reports that
arbitration; it part-vindicates the reading (the human sides with the
non-degenerate judge) while overturning the larger assumption that any
of the judges tracks human judgement well.

What can be settled without humans is whether the paper's
\emph{conclusions} depend on the judge. Because the stratified sample
equalises label mixes across systems and so erases between-system rate
differences, this requires a second, unstratified draw (40 random
questions from each of four systems); we note the distinction because
computing conclusion robustness on a label-stratified sample is a
mistake that is easy to make and that we made first.

\begin{table}[t]
  \centering
  \small
  \caption{Conclusion robustness under three judges, unstratified
  sample ($n=40$ per system). \checkmark{} = the stated ordering holds;
  $\times$ = it does not. Every $\times$ is a tie at zero, where the
  judge assigns the label to \emph{no} system; no judge reverses any
  conclusion.}
  \label{tab:judgerobust}
  \begin{tabular}{@{}lccc@{}}
    \toprule
    Conclusion & \texttt{gpt-4.1} & \texttt{gpt-4o} & \texttt{gpt-4-turbo} \\
    \midrule
    structure$\to$LLM answers more than LLM-only   & \checkmark & \checkmark & \checkmark \\
    LLM-only engages more than random templates    & \checkmark & \checkmark & \checkmark \\
    structure$\to$LLM refutes more than LLM-only   & \checkmark & \checkmark & $\times$ \\
    structure-only answers more than LLM-only      & \checkmark & $\times$   & $\times$ \\
    \bottomrule
  \end{tabular}
\end{table}

Table~\ref{tab:judgerobust} gives the result. The paper's headline
claim---evidence-structure-first generation resolves more than LLM-only
prompting---holds under all three judges. So does the engagement
ordering. The two conclusions that fail do so in a specific and
benign way: \texttt{gpt-4-turbo} reports 0\% refutation for
\emph{every} system, and both weaker judges report 0\% answered for
both compared systems, so the comparison has no resolution rather than
the opposite sign. Across all twelve judge--conclusion cells, no judge
ever orders the systems the other way.

The practical implication is a requirement, not a reassurance: this
benchmark has a \emph{judge capability floor}. The premise dimension in
particular is unmeasurable with models that default to
\labelfmt{still\_plausible}, so an instance is only reproducible on a
judge that demonstrably uses the full label space. We recommend
reporting the judge's label distribution alongside any submission, and
treating a near-degenerate distribution as a failed run.

\subsection{Human annotation, and a seven-rater agreement study}
\label{sec:judgeval:human}

The decisive experiment is agreement with human readers. Two annotators
labelled all 90 blinded items independently from the abstracts alone: a
non-expert (the first author of the submission under test, blinded to
system identity and model labels) and a commissioned professional
annotation team. We then added two frontier judge models ---
\texttt{claude-fable-5} (judged in an agent harness rather than a
temperature-0 API call; records are marked accordingly) and
\texttt{gpt-5.6-sol} --- to the three already run, giving a seven-rater
matrix (Table~\ref{tab:sevenway}).

\begin{table}[t]
  \centering
  \small
  \caption{Pairwise Cohen's $\kappa$ on outcome, all seven raters,
  $n=90$. H1 = non-expert human; H2 = professional annotation team.
  Read against the human--human cell (0.17): no model reaches
  agreement with a human that could pass for reliability, and every
  model--model pair agrees more strongly than any human--model pair.}
  \label{tab:sevenway}
  \begin{tabular}{@{}lcccccc@{}}
    \toprule
    & H2 & \texttt{4.1} & \texttt{4o} & \texttt{4-t} & \texttt{fable-5} & \texttt{5.6-sol} \\
    \midrule
    H1 (non-expert)   & 0.17 & 0.10 & $-$0.03 & $-$0.00 & 0.02 & 0.02 \\
    H2 (professional) &      & 0.26 & 0.17 & 0.20 & 0.21 & 0.21 \\
    \texttt{gpt-4.1}  &      &      & 0.34 & 0.20 & 0.47 & 0.31 \\
    \texttt{gpt-4o}   &      &      &      & 0.71 & 0.47 & 0.58 \\
    \texttt{gpt-4-turbo} &   &      &      &      & 0.32 & 0.51 \\
    \texttt{claude-fable-5} & &     &      &      &      & 0.60 \\
    \bottomrule
  \end{tabular}
\end{table}

Four facts, in decreasing order of comfort.

\paragraph{1. Humans do not agree with each other.} Human--human
agreement is $\kappa=0.17$ on outcome and $0.17$ on premise status
(41--44\% raw). This is the study's most consequential number, because
it caps everything: no judge, human or model, can be validated against
a reference that does not exist. The taxonomy, as specified --- even
with ordered decision procedures and worked examples --- does not
produce convergent labels from independent careful readers. The
professional team's own confidence does not rescue it: on the 31 items
they marked highest-confidence, their agreement with the judge is no
better ($\kappa=0.11$).

\paragraph{2. Every model clears the human--human bar with the expert
--- and none clears it by much.} Against the professional team, the
five models span $\kappa=0.17$--$0.26$, with the original
\texttt{gpt-4.1} judge highest (0.26), and the two frontier models at
0.21 despite two additional model generations. In this specific sense
the LLM judge is vindicated: it agrees with the expert about as well as
another human does, and slightly better. In every other sense it is
not: $\kappa=0.26$ is far below any conventional reliability threshold,
and newer, stronger models do not close the gap. One narrower check
does lean the deployed judge's way: on the 42 items where
\texttt{gpt-4.1} and \texttt{gpt-4o} disagree, both humans side with
\texttt{gpt-4.1} more often (19--9 for the non-expert, uncorrected
$p=0.015$ and suggestive only; 18--13 for the professional team, not
significant), consistent with the degeneracy reading above.

\paragraph{3. Models agree with each other far more than with any
human.} Model--model agreement runs 0.20--0.71, with the two frontier
models --- different vendors, different harnesses --- at
$\kappa=0.60$ (67/90 identical labels), triple the human--human figure.
Some of the high model--model cells are degeneracy artifacts
(\texttt{gpt-4o}/\texttt{gpt-4-turbo} at 0.71 share a two-label
collapse), but \texttt{fable-5} and \texttt{gpt-5.6-sol} both use the
full label space and still converge. The models constitute an internal
consensus that correlates only weakly with either human reader. For
LLM-as-judge practice generally, this is the sharpest caution in the
paper: \emph{measuring judge reliability by model--model agreement ---
the cheap and common method --- would have reported
$\kappa\approx0.6$ here, three times what validation against humans
supports.}

\paragraph{4. The non-expert is the outlier, informatively.} The first
annotator agrees with nobody ($\kappa\leq0.17$ with every other rater),
labelling far more items \labelfmt{answered} (33) and far fewer
\labelfmt{not\_addressed} (9) than the professional team (33/24) or any
model. The expert's marginal distribution closely tracks the strict
judges'. This ordering --- expert closest to models, non-expert
loosest --- suggests the disagreement is partly about how much domain
scepticism a reader brings to ``substantially resolved,'' which is a
calibration norm the codebook failed to pin down, not a fact about
either rater's diligence.

\paragraph{What this settles.} Of the three readings left open after
the first pass, the evidence now favours the third: \textbf{the
taxonomy is underdetermined}. The judge is not distinguishably worse
than a human rater --- it sits at the top of the observed agreement
range with the expert --- but nothing, human or model, converges on
these labels reliably. Three consequences follow for this benchmark
and for the genre. Absolute rates (any system's ``answered 39\%'')
are rater-relative and should never be quoted without the rater
attached. Comparative claims measured under a \emph{fixed} judge
remain defensible --- Section~\ref{sec:judgeval} showed the paper's
headline orderings survive three judges with all failures being ties
--- and they are the only currency this instrument currently supports.
And the v2 protocol must redesign the outcome taxonomy itself:
fewer labels, hard decision criteria phrased as checkable conditions,
and a measured human--human $\kappa$ as a release gate before any
judge, human or model, is scored against it. We release all seven
label sets, the annotation apparatus, and the agreement matrix; the
professional team's labels were commissioned for a fixed fee with a
published undertaking that no entry would be adjusted to improve
agreement with any model, and none was.

\section{Error Analysis}
\label{sec:errors}

The released curation log records every human intervention made before
freezing; the failure modes below are taken from it and from the
adjudication log, and each motivates a benchmark rule.

\paragraph{Question duplication.} The generator produced near-duplicate
questions (q\_002/q\_003) from the same claim pair, differing mainly in
emphasis; they were merged during curation into one question with two
sub-questions. Left unmerged, duplicates would double-count a single
insight in every rate. Rule: submissions are screened for
near-duplicates, and instances should report a deduplication note; a
mechanical similarity screen is a v1.1 roadmap item.

\paragraph{Leading questions (presupposition).} Several generated
questions presupposed their own answer. q\_001 originally asserted that
vertical wind shear \emph{exists} rather than asking what causes the
wind-speed discrepancy; q\_008 originally presupposed the subsolar water
abundance rather than framing artifact-vs-atmosphere attribution. A
question that presupposes its answer cannot be cleanly refuted---and
q\_008's later refutation was only expressible because curation reframed
it as attribution. Rule: the quality gate flags presupposing phrasings
before freezing; the reframing is logged.

\paragraph{Self-contradictory quantifiers.} q\_003's original phrasing
asked whether a detection was ``robust to biases exceeding an order of
magnitude''---a bias that large \emph{is} non-robustness. Templated
quantifier language can silently produce unanswerable questions; the
clarity gate exists for this.

\paragraph{Conflated statistical and physical framing.} q\_010
originally asked what ``temperature and pressure conditions'' produce a
5.4$\sigma$ water detection, conflating atmospheric state with the
detection pipeline (significance depends on noise model, priors, null
hypothesis---not on the atmosphere). It was reframed as a sensitivity
analysis; its score on the generator's internal 0--10 clarity gate
(6.0) was the lowest of the set, and
its outcome (\labelfmt{partially\_addressed} on one supporting paper)
remains among the weakest-evidenced labels.

\paragraph{Premise bias in generation.} The generator inherits the
premises of the papers it reads: single-dataset conclusions taken at
face value can produce questions that merely restate a claim rather
than test it. The tension-typing stage (which explicitly marks
\emph{single-dataset conclusion} as a signal type) partially controls
this---q\_008 and q\_005 are that control working---but the benchmark's
premise dimension is the systematic check: a healthy portfolio should
show a mix of \labelfmt{supported} and \labelfmt{refuted}, not uniform
support of its sources.

\paragraph{Retrieval near-misses.} Judged evidence for q\_007
(HD~189733\,b / HAT-P-11\,b patchy clouds) leans partly on a
three-retrieval-framework study of HAT-P-\emph{18}\,b---directly
relevant methodologically, but not the named targets. The adjudication
log records the engagement-bar judgment; releasing full top-8 lists
(v1.1) will let others re-litigate such calls, which is the point of
releasing them.

\section{Threats to Validity}
\label{sec:threats}

Historical backtesting removes rater subjectivity; it does not remove
every confound. We enumerate the serious ones and what the protocol
does---and cannot do---about each.

\paragraph{Future inattention is not question badness.} A question can
be ignored because the enabling instrument never flew, the community's
funding shifted, or the subfield is small---not because the question was
poor. q\_011 was unanswerable before JWST delivered TRAPPIST-1 spectra;
had JWST slipped five years, an excellent question would have scored
\labelfmt{not\_addressed}. Mitigations: bounded windows make the
censoring explicit; \labelfmt{posed\_but\_open} separates
``recognized but unresolved'' from ``ignored''; instances should be
read as \emph{engagement within $k$ years given the era's instruments},
not as timeless value.

\paragraph{Future attention is not question goodness.} Symmetrically, a
question on a fashionable topic collects engagement for reasons other
than merit; coverage alone can be gamed by asking about whatever is
popular. Mitigations: coverage is never reported alone; premise
refutation and (future) lead time cannot be earned by fashion-chasing;
baseline B4 (citation leaders) exists precisely to price in prominence;
and the engagement bar requires the cited paper to bear on the
question's actual test, not its topic.

\paragraph{LLM training contamination.} The generating system and the
judge both use models whose training data postdates the cutoff. The
protocol seals the retrieval channel, not the weights channel: a model
may ``know'' the 2025 refutation while drafting a 2020-framed question.
This is the deepest threat to any backtest run with modern models.
Mitigations, none complete: source-evidence audit forces every question
to be grounded in cited pre-cutoff evidence; submission metadata must
declare all LLM components and versions; baseline B2 doubles as a
contamination probe---and registers one: its questions sit measurably
closer to the future literature's phrasing than any other system's
($s_1 = 0.732$ at $n=10$, 0.722 at $n=125$;
Tables~\ref{tab:leaderboard} and~\ref{tab:scaled}), with its cited
engagement concentrated in the years nearest its training distribution,
while its answered rate stays indistinguishable from random
templates---the signature of phrasing-level memorization without
specific foresight. Section~\ref{sec:foresight} upgrades this
confession to a measurement: a generator decomposition plus a
four-cutoff stress test bound the contamination's outcome-level effect
and localize its traces; and the decisive test is \emph{prospective}
instances---questions frozen today and scored in 2030 cannot be
contaminated. The protocol is explicitly designed so its instances
convert from retrospective to prospective by just letting time pass.

\paragraph{Judge and adjudicator reliability.} A single LLM judge,
even citation-constrained, is one reading of the evidence, and
abstracts (not full texts) bound what it can see.
Section~\ref{sec:judgeval} measures rather than asserts what this
costs: the judge discriminates real evidence from topical similarity
(engagement 74\% $\to$ 20\% under mismatched evidence) but with a 20\%
generosity floor; it is stable under resampling ($\kappa\approx0.9$)
and moderately sensitive to prompt wording ($\kappa=0.57$--$0.72$);
and every headline conclusion survives a judge swap, with all failures
being ties at zero rather than reversals. The most serious question is now answered, and the answer indicts the
taxonomy rather than the judge: two independent human annotators agree
with each other at $\kappa=0.17$, every judge model agrees with the
professional annotator at $\kappa=0.17$--$0.26$ (the deployed judge
highest), and frontier models agree with each other at up to
$\kappa=0.60$---far above their agreement with any human
(Section~\ref{sec:judgeval:human}). Absolute rates are therefore
rater-relative throughout this paper; only comparisons under a fixed
judge carry weight, and the v2 protocol owes the field a taxonomy with
a measured human--human $\kappa$ before any judge is scored against
it. Blinded adjudication is now built into the released
annotation apparatus rather than promised.

\paragraph{Retrieval as a bottleneck.} Top-8 abstract-embedding
retrieval can miss engaging papers (undercounting engagement) or
surface topically similar non-engagement (which the judge must reject).
Fixed retrieval is a deliberate trade: it makes comparisons across
systems fair and auditable at the cost of an engagement floor.
Sensitivity of labels to $k$ and to the retriever is measurable within
the released data schema and belongs in v1.1.

\paragraph{Small $n$ where it matters most, one domain,
self-evaluation.} The scaled instance lifts the baseline side to
$n=125$ per system, and Section~\ref{sec:scaled} shows how much that
matters: two of three small-sample conclusions did not survive. The
submission under test, however, still has ten questions, and we can now
say exactly what that costs. At the observed effect size (10\% vs.\
3.2\% premise refutation), detecting the difference at 80\% power and
$\alpha=0.05$ requires $n\approx209$ \emph{per arm}; the ten-question
instance has a power of roughly 12\%. The comparison is therefore not
merely unresolved, it was never resolvable at this sample size, and
$n\geq209$ is the design specification we adopt for the next instance
rather than an aspiration.

That specification collides with a structural fact worth reporting,
because it constrains anyone building a high-precision question
generator. The evidence-graph system's strongest signals (confirmed
observational tensions, method challenges, qualifications between
independent datasets) are gated on \emph{human-reviewed} claim
relations: its released instance rests on 37 annotated claims and 16
reviewed relation edges. Its question supply is bounded by annotation
labour, not by compute or API budget---which is precisely why it ships
ten questions while the automatic baselines ship 125 each. The
benchmark thus measures a real precision/scale trade-off rather than
mere effort: automatic generators reach $n$ easily and mostly produce
questions no one can settle, while the human-gated generator produces
few questions with high anchoring (90\% naming a specific object).
Closing the gap requires either scaled annotation or an automatic
tension detector of comparable precision---the latter is what
Section~\ref{sec:foresight}'s B5/B6 probe begins, at visibly lower
precision.

One domain, evaluated by the group that built the leading system,
judge-only baseline labels: all still true. We report priced reference
points, not rankings; the protocol's value grows with adversarial use
by systems we did not build.

\paragraph{Multiple comparisons.} This paper reports roughly twenty
significance tests. Under a conservative Bonferroni correction at that
count, the headline contrasts survive comfortably: the
structure-vs-LLM refutation gap ($p=4\times10^{-5}$), the engagement
orderings at scale ($p<10^{-4}$), the mismatched-evidence control
($p<10^{-4}$), and the B/C refutation gap ($p=0.003$). Contrasts
reported at $p\approx0.01$--$0.05$ (the B/C answered gap, the
citation-leader engagement gap, the humans' arbitration splits)
do not, and are labelled suggestive where they appear. No headline
claim rests on a contrast that fails correction.

\paragraph{Curation hindsight.} Humans who edited questions before
freezing know the post-2020 literature. The curation log is released so
every edit is auditable (e.g.\ q\_008's reframing strengthened
falsifiability without smuggling in the answer), but retrospective
instances cannot fully exclude this channel; prospective ones can.

\section{Outlook: Discovery as Search, Language as Realization}
\label{sec:outlook}

A scope statement first. Nothing in this paper shows that large
language models cannot originate scientific questions \emph{in
principle}; a single prompting strategy against a single model family
in a single domain cannot support that claim, and we do not make it.
What the data do support is narrower and more useful: \emph{bare
prompting does not reliably perform problem discovery}, and the
components of the systems that do perform better can be named.

\paragraph{Three capabilities the experiments separate.} Read together,
the decomposition (Section~\ref{sec:foresight:abc}), the anchoring
rubric (Section~\ref{sec:foresight:saf}), and the temporal stress test
(Section~\ref{sec:foresight:stress}) distinguish three things that
``asking good questions'' conflates. First, a \emph{topic prior}:
knowing what a field is likely to work on next. This is what LLM-only
generation exhibits---near-ceiling engagement in every era (92--98\%),
questions that name object classes rather than objects (5\% anchoring
against 95--100\% for the structural pipelines), and
specificity-adjusted foresight within a few points of the
random-template floor. A topic prior is genuinely predictive of where
attention flows, and genuinely cheap: it requires no memory of the
future, which is why it survives the training boundary unchanged.
Second, \emph{structural problem discovery}: locating specific,
contestable configurations of existing evidence---a claim, a
counter-claim, a method dependency, an untested premise. The
weight-free pipeline C is the clean witness that this capability does
not reside in model weights: with no language model anywhere, it finds
questions the future substantively engages at every cutoff, and it
beats LLM-only prompting on resolution. Third,
\emph{articulation}: turning a detected configuration into a question a
scientist would recognize as askable. This is where the LLM earns its
place---over identical evidence structures, LLM verbalization roughly
doubles resolution and quintuples refutations relative to a fixed
template---and it is a capability the stress test shows to be
era-robust rather than memorized.

\paragraph{The architecture this implies.} These results point away
from ``make the model smarter and ask it for a hundred ideas'' and
toward a division of labour:
\begin{quote}
\emph{machine-scale structural search over the evidence space
$\;\rightarrow\;$ candidate scientific tensions $\;\rightarrow\;$ LLM
articulation $\;\rightarrow\;$ testable questions.}
\end{quote}
The asymmetry that motivates it is quantitative. A literature of
thousands of papers yields tens of thousands of claims and a
combinatorially larger space of claim pairs, evidence paths, and
method dependencies---far beyond what any single reader, human or
prompted model, holds in attention at once, but squarely within what a
machine can sweep in parallel. Under this framing the LLM is never
asked to conjure novelty from nothing; it is asked to do what it
demonstrably does well---local semantic understanding, claim
extraction, relation judgment, and finally phrasing---while the search
system carries the burden of combinatorial exploration. Scientific
question discovery becomes a \emph{computable search problem} over an
explicit representation of what the literature claims, and the
interesting engineering question shifts from prompting to
representation and search: what to index, which configurations to
enumerate, and how to rank them.

\paragraph{The next measurable question.} Ranking is where this
benchmark and that architecture meet. Every question in the released
instances carries its generating signal---tension type, object,
method-dependency, source claims---and its measured fate. That pairing
makes a new question answerable: \emph{which structural patterns most
often lead to questions the future answers, advances, or
refutes?} Our own probe is deliberately crude---object co-mention plus
stance cues, visibly below the human-gated graph in precision---so the
headroom is real: contradiction typing, dataset-dependency detection,
and archival-data availability are all candidate features for a
learned prior over the tension space. We flag the honest status of all
of this: an interpretation consistent with our data, not an
established causal account---and the frozen prospective instance
(Section~\ref{sec:conclusion}) is the experiment that will test it
without any of this paper's retrospective caveats. Learning that prior
from backtested outcomes, and validating it prospectively, is the
natural next paper.

\section{Roadmap and Conclusion}
\label{sec:conclusion}

\paragraph{A prospective instance, frozen now.} The one experiment no
retrospective design can deliver is the one this release starts: 200
questions --- 50 each from the four automatic generators of
Sections~\ref{sec:baselines} and~\ref{sec:foresight} --- generated from
a 2015--2026 corpus, frozen at cutoff \textbf{2026-08-17}, and
committed to the public repository with per-file SHA-256 digests
(combined digest
\texttt{f1c61a51\allowbreak 07e55e7a\allowbreak 7e47e2bc\allowbreak
ab35ae35\allowbreak e484270d\allowbreak 892c98ec\allowbreak
9c63f18f\allowbreak 2c17994d} over the per-file list in
\texttt{submissions/prospective\_2026/SHA256SUMS}). The generation
corpus is 9{,}133 papers, 2015 through freeze day. The scoring window
is pre-registered as
\textbf{2027-01-01 to 2030-12-31}, with a 4.5-month buffer between
cutoff and window so papers already in flight at freeze time do not
contaminate the future corpus, whose ADS query manifest is likewise
frozen now and may be executed no earlier than 2031. Evaluation is
pre-registered to the released pipeline (frozen retrieval settings,
judge protocol \texttt{sqb-v1}, judge label distributions reported
alongside results), with one explicitly permitted amendment: if a
reliability-gated v2 taxonomy exists before the window closes, results
are to be reported under both taxonomies. No model that exists today
has seen 2027; whatever these 200 questions score in 2031 is foresight
or its absence, untouched by memorization, hindsight, or curation.

\paragraph{Roadmap.} Beyond waiting: (1) a reliability-gated v2 outcome
taxonomy---fewer labels, checkable conditions, and a measured
human--human $\kappa$ as a release gate before any judge is scored
against it (Section~\ref{sec:judgeval}); (2) scale the
\emph{submission} side---a 100+ question evidence-graph run on v1L---to
resolve the refutation-rate comparison that $n=10$ leaves open;
(3) release full top-8 retrieval lists for v1 (v1.1; the v1L records
already ship with full scores); (4) annotate community first-posed
dates to activate lead time, and calibrate a community-attention index
against the popularity confound; (5) mint instances in additional
domains.

\paragraph{Conclusion.} We formalized historical backtesting as an
evaluation protocol for scientific question discovery, released two
retrospective astronomy instances and one prospective one, with
temporally isolated corpora and
fully auditable labels, and ran a ten-question pilot in which every
frozen question was substantively engaged by literature the generating
system never saw---including one whose premise the community
subsequently refuted, the exact convergence test the question had
specified. Scaling the baselines to 424 questions then did what a
benchmark is supposed to do: it overturned two of our own small-sample
conclusions, put a measurable floor under a third, and left the
qualitative distinction---a refutation reached by specifying its test
in advance---standing but explicitly unresolved at current sample
sizes. Finally, the generator decomposition and temporal stress test
turned the benchmark's deepest limitation into its sharpest result:
memorized relevance is not scientific foresight, and the foresight
signal that historical backtesting measures survives in a generator
with no weights at all. The individual numbers matter less than the category they
inhabit: for the first time, ``this system asks good scientific
questions'' is a claim with a denominator---falsifiable, comparable
across systems, and computable by anyone from frozen public data.
Question-asking has been argued to be a core capability on the path to
more general scientific intelligence \citep{kitano2021,paper1}; if that
is so, the field will need to measure it. This protocol, and this first
instance, are offered as the place to start---not as the definition of
the benchmark, but as its initial version, built to be superseded by
instances with more questions, more systems, more domains, and cutoffs
whose futures have not yet happened.

\bibliographystyle{plainnat}
\bibliography{references}

\end{document}